# ORIGIN OF ECHO-ENABLED HARMONIC GENERATION*

Vladimir N Litvinenko, Stony Brook, USA

*Abstract*

In this paper I present overview of two preprints that were published at Budker Institute of Nuclear Physics in 1980 [1,2]. These original publications describing harmonic generation technique which has the same foundation as Echo-Enabled Harmonic Generation (EEHG). They recently become available online [3], but they are still unknown to FEL community. This paper is an attempt to current this omission.

## INTRODUCTION

When I first read about EEHG [4], it vaguely reminded me of theory developed at Novosibirsk Institute of Nuclear Physics (BINP) in late 1970s. But I was not 100% sure that my memory of decades-old events is correct. Recently, the BINP made their preprint available on the web and I was able to confirm that technique proposed by I.G. Idrisov and V.N. Pakin [1,2] based on the similar principles to those described in [4]. In this presentation I would like to briefly review these original papers and give credit to early inventors of this innovative high-harmonic generation technique.

These papers- see Figure 1 - were written in Russian in form of preprints and were not available outside the Iron Curtain – a typical totalitarian method of preventing the exchange of information. Hence, in general, only people from BINP had direct exposure to these findings.

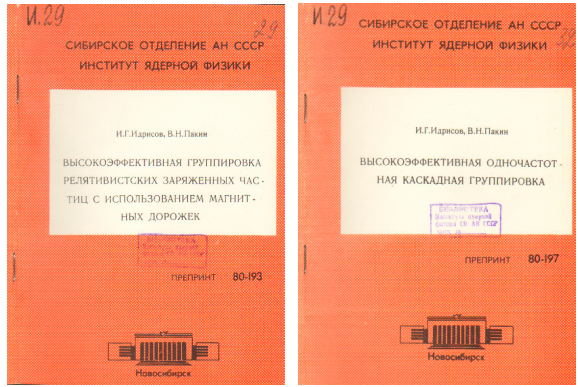

Figure 1: Cover pages of the two preprints.

The goal of this short paper is to give historic perspective on EEHG and to provide FEL community with references to these original papers. I attempted to translate snippets of original texts and use as many as possible of original illustrations. Unfortunately, quality of some illustrations is poor.

* Work supported by DE-SC0020375 grant from the Office of High Energy Physics, US Department of Energy.

## CONTENT OF TWO PREPRINTS

It is impossible to include all text, formulae, figures, and tables from two preprint in this 4-page paper. When it is necessary, I will put in brackets references to these preprints. For example, Fig. {1,3} or eq. {2.5} will mean Figure3 in reference [1] and equation (5) in reference [2]. correspondingly. Similarly, Section {2, 2.3} means section 2.3 in reference [2]/.

*Abstracts*

{1,1} We demonstrate possibility of high efficiency bunching of relativistic beams using magnetic compressors. We show that amplitude of the current in in first harmonic is possible to increase to 1.9 $I_0$ and the current in 10$^{th}$ harmonic to 1.8 $I_0$ (where $I_0$ is the current in un-bunched beam), or even higher….. High efficiency in amplitudes of high harmonics is achieved using strong compression (*to be exact – over-compression using large R56 - VL*) at initial stages and very strong (increasing) energy modulation in the later stages of the bunching.

{2,1} We show in this paper that using strong compression (e.g. over-compression) in first compressors, followed by increasing amplitude of energy modulation in the following steps of cascade bunching system allows to increase efficiency of the modulation to 1.76 $I_0$, 1.94 $I_0$ and 1.98 $I_0$ (where $I_0$ is the current in un-bunched beam) using two, three and four cascades, correspondingly.

*{2.2} Kinematic approach*

*{2,2.1} Cascade Bunching*

Fig. 1.1 shows a typical schematic of cascade bunching.

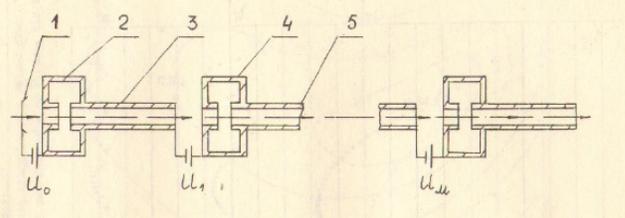

Figure 2. {2,1}. Schematic of multi-stage bunching

A continuous beam with current $I_o$ from the cathode (1) is accelerated to energy $eU_o$ and passes through first RF cavity (2), where its energy(velocity) modulated by RF voltage $U_{m1}$. The drift (3) turns energy modulation into density (current) modulation….. This process is repeated with $U_1$, $U_2$, … representing modulation at each stage.

*{2,2.2} Idealized equations for phase motion. Efficiency of harmonic generation .*

For simplification, let's consider a reference particle which does passing all cavities at zero-crossings:

$$U_1 = U_2 = \ldots = 0;$$

... We assume that relative energy modulation is small that second order effects can be neglected, and that depth of energy modulation is increasing from stage to stage. ...

These assumptions allow us to write following set of equations:

$$\varphi_1 = \varphi - \chi_1 \cdot \sin\varphi \quad (1)\ \{2.1\}$$

with compression parameter

$$\chi_1 = \frac{\alpha_1 \theta_1}{2}, \quad (2)\ \{2.2\}$$

where $\theta_1$ is drift parameter* and

$$\alpha_1 = \frac{U_1}{U_o};\ \alpha_i = \frac{U_i}{U_o};\ \alpha_1 \ll \alpha_2 \ll \ldots \ll 1 \quad (3)\ \{2.3\}$$

being the depth energy modulation in first section†. Then at the exit of 2 stages we have {2.4}:

$$\varphi_2 = \varphi_1 - \chi_2 \cdot \sin\varphi_1 = (\varphi - \chi_1 \cdot \sin\varphi) - \chi_2 \cdot \sin(\varphi - \chi_1 \cdot \sin\varphi) \quad (4)$$

By induction, at $k^{th}$ stage, we can write eq. {2,5}:

$$\varphi_k = \varphi_{k-1} - \chi_k \cdot \sin\varphi_{k-1} = (\varphi_{k-2} - \chi_{k-1} \cdot \sin\varphi_{k-2}) - \chi_k \cdot \sin(\varphi_{k-2} - \chi_{k-1} \cdot \sin\varphi_{k-2}) = \ldots \quad (5)$$

We expand beam current at the exit of $k^{th}$ stage {2,6}:

$$i_k = I_o \left(1 + 2\sum_{n=1}^{\infty} f_{n,k}(\chi_1, \chi_2 \ldots \varphi_k) \cos n\varphi_k \right) \quad (6)$$

where Fourier coefficients $f_{n,k}$ represents amplitudes of $n^{th}$ harmonic in beam current units of $2I_o$, and therefore, the maximum possible efficiency of generating $n^{th}$ harmonics {2,7}:

$$\eta(n,k) = f_{n,k} = \frac{1}{\pi}\int_0^\pi \cos(\varphi_k(\chi_1, \chi_2 \ldots \varphi)) d\varphi \quad (7)$$

Integral (6) can be analytically evaluated for $k=1$, but for arbitrary $k$ should be evaluated numerically.

### {2,2.3} Results of numerical simulations.

To find maximum possible efficiency (7) it is necessary to find extremum of $f_{n,k}(\chi_1, \chi_2 \ldots \chi_k)$. We used Gauss-Seidel method described in D.J. Wild, Extremum Methods, Nauka, Moscow (1967). Results for systems with two and three stages are shown in Tables I and II. It is possible to notice that parameters $\chi$ approaching

---

* Since focus of these papers was on non-relativistic devices, authors used not a modern notation such as $R_{56} = \theta_1/2$.

† In modern language relevant to EEHG $\alpha_1 = \frac{\delta\gamma_1}{\gamma_o}$

---

constant values for high harmonic numbers, which we could call "asymptotic compression parameters".

Table I {2,I}. Maximum efficiency for system with two stages

| n | $\eta_e(n,2)_{макс}$ | $\chi_1$ | $\chi_2$ |
|---|---|---|---|
| 1 | 0,879 | 3,11 | 1,51 |
| 2 | 0,842 | 2,82 | 1,33 |
| 3 | 0,819 | 2,65 | 1,25 |
| 5 | 0,790 | 2,47 | 1,18 |
| 10 | 0,749 | 2,24 | 1,11 |

Table II {2,II}. Maximum efficiency for system with three stages

| n | $\eta_e(n,3)_{макс}$ | $\chi_1$ | $\chi_2$ | $\chi_3$ |
|---|---|---|---|---|
| 1 | 0,971 | 4,08 | 2,75 | 1,28 |
| 2 | 0,962 | 3,97 | 2,50 | 1,20 |
| 10 | 0,937 | 3,60 | 1,92 | 1,05 |

We calculated efficiencies of first 50 harmonics for compression parameters $\chi$ close to asymptotic values

K = 1   $\chi_1 = 1,08$
K = 2   $\chi_1 = 2,08$   $\chi_2 = 1,08$
K = 3   $\chi_1 = 3,60$   $\chi_2 = 1,92$   $\chi_3 = 1,05$

and summarized them in Table III.

Table III {2,II}. Efficiencies of harmonic generation in systems with one, two and three stages

| n | $\eta_e(n,1)$ | $\eta_e(n,2)$ | $\eta_e(n,3)$ |
|---|---|---|---|
| 1 | 0,465 | 0,823 | 0,959 |
| 3 | 0,349 | 0,778 | 0,947 |
| 10 | 0,269 | 0,740 | 0,937 |
| 30 | 0,213 | 0,672 | 0,914 |
| 50 | 0,170 | 0,570 | 0,878 |

This table shows that $\eta(n,k)$ remains large for n>10 for two stage, and especially for three stage system. It means that current distribution approaching an ideal, δ-function-like, distribution. Figure 3 {2} illustrates phases at the exits of the first, second and third stages. ..... Figures 4 and 5 show that area of efficient harmonic generation is reduces for higher harmonics… Further numerical studies, not included in this preprint, showed that area with high efficiency of harmonic generation increases when number of cascades is increased.

*{2,2.3} Discussions.*

Our studies showed that large compression parameters are needed for high efficiency of harmonic generation in multistage system ….. Our simulations showed that inclusion of modest velocity spreads does not significantly affect efficiency of harmonic generation with $n \leq 10$…

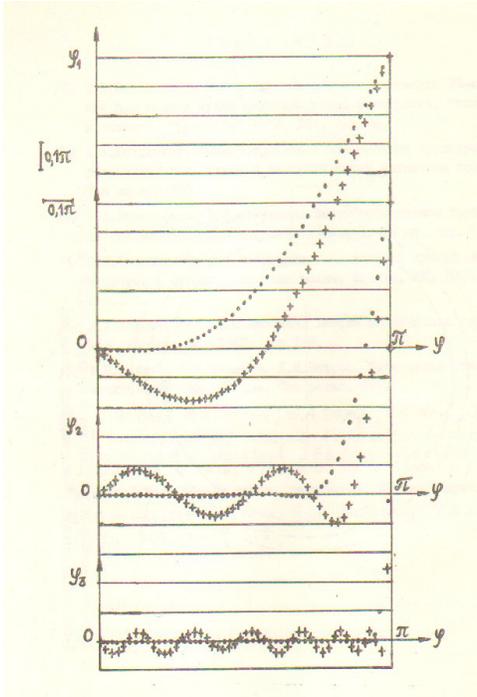

Figure 3 {1,2}. Dependence of phases at the exits of the first, second and third stages on that at the entrance of the system. Crosses show them for compression parameters optimal for first (fundamental0 harmonic. Dots are compression parameters close to asymptotic values. 50 traces were used.

Figures 4 and 5 show contour plots if the efficiencies for generating first (fundamental) and fifth harmonics in tow stage system.

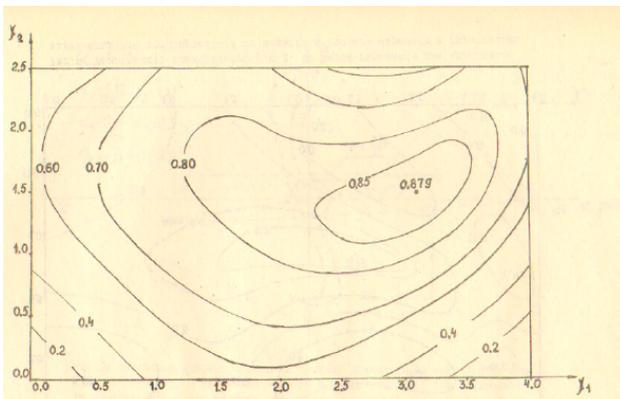

Figure 4 {2,3}. Efficiency contour plot for first harmonic in two stage system as function of $\chi_{1,2}$

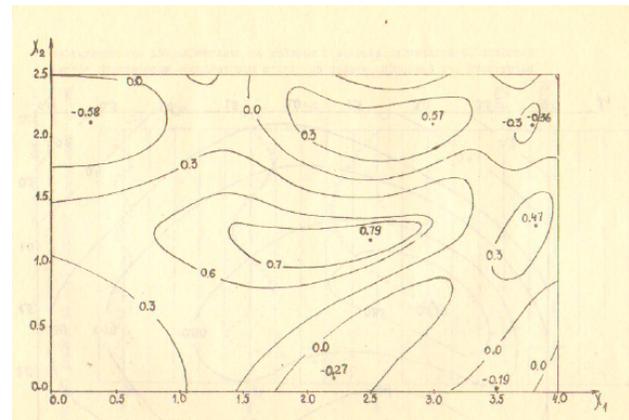

Figure 5 {2,4}. Efficiency contour plot for fifth harmonic in two stage system as function of $\chi_{1,2}$.

Figure 6 show an example of the phase space evolution in two-stage system with $\chi_1 = 3.12$ and $\chi_2 = 1.51$ and $\alpha_2/\alpha_1 = 5$ …..

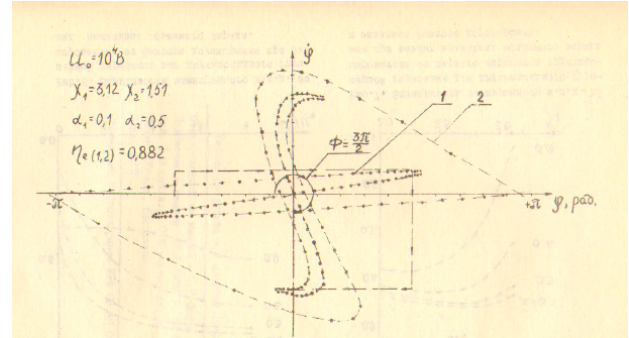

Figure 6 {2,5}. Location of test particles in the longitudinal phase space $(\varphi, \dot\varphi)$. Curve 1 – at the exits of the stage one; Curve 2 – at the exit of stage 2.

101 test particles were used to simulate evolution in this two-stage system. Arrows in the figure show phase trajectories of the particles with maximum perturbation in the first stage of the system. One can see that location of these particles rotated in the phase-space by $\Phi = 3\pi/2$ with respect to initial position The overall particle flow resembles whirling about the origin with strong phase compression towards the center.…. Increasing number of stages increases number of swirls with critical trajectories rotating by $5\pi/2, 7\pi/2$, etc.

*{1,2.4} Discussions of results.*

*Necessity of applying strong energy modulation and compression parameters to achieve effective harmonic generation, $f_{n,k}$, is related a sine-wave modulation of the energy… Hence, the goal of effective bunching is to compress as many particles as possible (locate in interval of initial phases of $\pm\pi/2$) close to the origin…*

Figure (7) shows step by step evolution of particles in a two-stage system with large ratio of energy modulation in the first and the second sections: $\alpha_2/\alpha_1 = 9$.

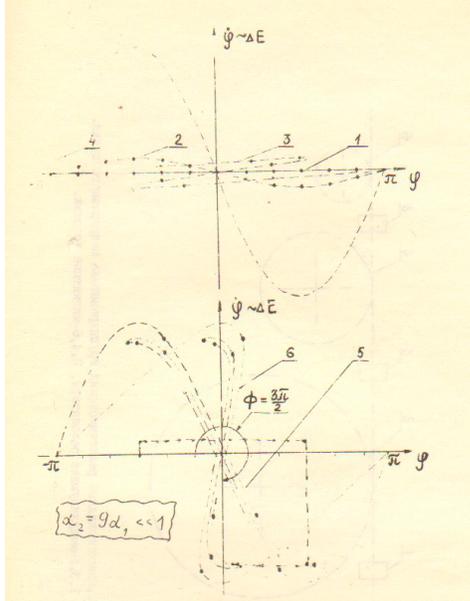

Figure 7 {1,5}. Particle's evolution in the longitudinal phase space for a two-stage system. Curves in these plots are as follows: 1 – initial particle's positions; 2 – positions after energy modulation in first section; 3- positions after the first drift; 4- energy modulation in the second section; 5 – positions after energy modulation in the second section; 6 - positions after second drift.

*{2.4} Conclusions.*

Our investigations showed that use of multi-stage systems with high compression parameters provides for high efficiency bunching and harmonic generation.

*This concludes review of these two preprints. Authors also discussed effects of space charge, which I omitted for compactness and relevance to EEHG.*

## DISCUSSIONS AND CONCLUSION

It is rather obvious that focus of two preprints was on bunching of relatively low energy beams used in klystrons, even though relativistic effects were considered in Ref. [1]. Vinokurov and Skrinsky clearly described both similarity and differences between beam dynamics in optical klystron [5], i.e. in an FEL, and its low energy cousin. Since then, it became apparent that many ideas, including harmonic generation, are applicable – albeit with careful formulation – to FELs. One of very important considerations in for FEL is clearly including well-known bunching suppressions caused by the uncorrelated energy spread [5]:

$$\rho_k \to \rho_k \exp\left(-\frac{1}{2}\left(\frac{kR_{56}\sigma_\gamma}{\gamma}\right)^2\right),$$

where is k-vector of the radiation.

Still, kinematics of the efficient high harmonic generation in FELs, known as EEHG [4], has the same origin as that described in the two preprints: (a) after first energy modulation bunch is over compressed to create filamentation in beam energy; (b) applying second energy modulation at the same frequency and optimum $R_{56}$, transfers filamentation into the longitudinal density modulation – see Fig. 8.

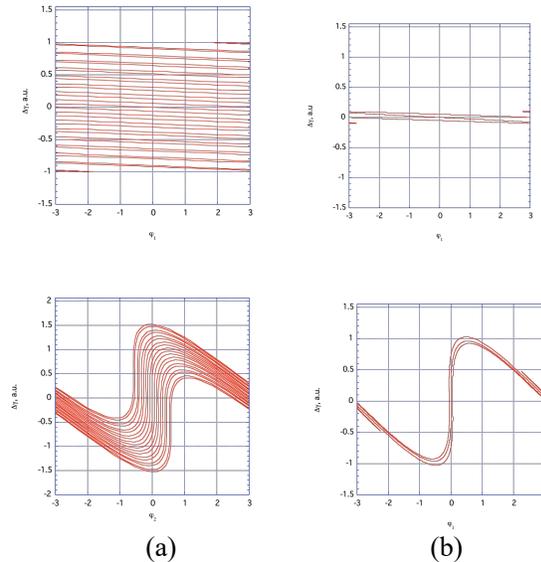

(a)          (b)

Figure 8. Phase space diagrams: top is after first section, and bottom – after the second section. Two "EEHG" schemes generate the same amplitude of harmonics but use different ratio between amplitude of energy modulation: (a) $\delta\gamma_1 = \delta\gamma_2$; (b) $\delta\gamma_1 = \delta\gamma_2/10$. Vertical scale is in units of $\delta\gamma_1$.

Finally, one of the interesting hints from these original preprints is that there is potential of furthering efficiency of high harmonic generation by using varying strengths of modulation in two, three, or even more, bunching stages. With modern computers these possibilities can be easily explored [6].